\documentstyle[epsf]{article}

\def\RR{{\mbox{\boldmath $R$}}}
\def\ss{{\mbox{\boldmath $s$}}}
\def\xxi{{\mbox{\boldmath $\xi$}}}
\def\sgn{{\rm sgn}}
\def\AGS{{\rm AGS}}

\title{Application of two-parameter dynamical replica theory
	to retrieval dynamics of associative memory with non-monotonic neurons}
\author{Toshiyuki Tanaka\dag, Shinsuke Kakiya\dag, and Masato Okada\ddag\\
\dag Department of Electronics and Information Engineering,\\
Tokyo Metropolitan University, Tokyo, Japan\\
\ddag Kawato Dynamic Brain Project,\\
Japan Science and Technology Corporation, Kyoto, Japan}
\begin{document}
\maketitle
\begin{abstract}
The two-parameter dynamical replica theory (2-DRT) is applied
to investigate retrieval properties of non-monotonic
associative memory,
a model which lacks thermodynamic potential functions.
2-DRT reproduces dynamical properties of the model quite well,
including the capacity and basin of attraction.
Superretrieval state is also discussed in the framework of 2-DRT.
The local stability condition of the superretrieval state is given,
which provides a better estimate of the region in which superretrieval
is observed experimentally than the self-consistent signal-to-noise
analysis (SCSNA) does.
\end{abstract}

\section{Introduction}
The Hopfield model has attracted interests of researchers
in various fields, and enormous amount of studies,
both numerical-experimental and theoretical,
have been carried out on it.
Of particular interest among them is the one on the model
with {\em non-monotonic} units\cite{Morita93}:
Whereas the conventional model uses, as the output function $f$ of a unit,
monotonic functions such as $f(x)=\tanh\beta x$ ($\beta>0$),
the model with non-monotonic units, or the non-monotonic model for short,
uses a non-monotonic function.
It has been reported that the non-monotonic model has
various nice properties as a model of associative memory,
such as enhancement of storage capacity,
enlargement of basins of attraction associated with
retrieval states.

In order to rigorously argue such properties of the non-monotonic model,
theoretical analyses are necessary.
However, attempts to analyze the non-monotonic model
are often faced with difficulty
because the non-monotonic model in general does not have
a Lyapunov function, which would be a powerful analytical tool
for characterizing the equilibrium as well as dynamical properties
of the model.
Thus, applicable theories are restricted to what have been
devised for analyzing
the conventional model and yet are independent of the functional form
of the output function $f$.
As for equilibrium analysis, the self-consistent signal-to-noise analysis
(SCSNA) has been applied to
the non-monotonic model\cite{SF93} and some interesting properties,
including the existence of the so-called superretrieval states,
have been found.
For retrieval dynamics, on the other hand,
currently no exact and tractable theory has been known
even for the conventional model:
The path integral formalism\cite{RSZ88}
and Gardner-Derrida-Mottishaw theory\cite{GDM87} (for ``zero-temperature,''
or $\beta\to+\infty$ case) are the exact theories
for the asynchronous (or Glauber) and synchronous (or Little)
dynamics, respectively,
but computation of dynamics based on each of them
is prohibitively difficult.
It has been generally believed\cite{RSZ88,HBFKS89,Okada95}
that any tractable theories
on retrieval dynamics necessarily incorporate some approximation.
For the conventional model with the synchronous
dynamics, Amari-Maginu theory\cite{AM88} (for zero-temperature case;
for extension to finite-temperature case, see \cite{NOz93}) has been proposed
as one of such theories, and Nishimori and Opri\c{s}\cite{NOp93} have applied
it to the non-monotonic case.
As one of tractable theories for the conventional model
with the asynchronous dynamics,
Coolen and Sherrington\cite{cs-l,cs} have proposed
the 2-parameter dynamical replica theory (2-DRT hereafter;
it is also sometimes called the Coolen-Sherrington (CS) theory).
Of course it is an approximate theory, as confirmed, for example,
by Ozeki and Nishimori\cite{on} and Tanaka and Osawa\cite{Tanaka98};
nevertheless, it describes retrieval dynamics of the conventional model
reasonably well.
As Ozeki and Nishimori\cite{on} have mentioned, formulation of 2-DRT
does not depend on the functional form of the output function $f$,
and therefore it is possible to apply it to the non-monotonic model.
Thus, the following question naturally arises:
{\em How well does 2-DRT describe retrieval dynamics
of the non-monotonic model?}
In this paper, we address this problem,
with emphasis placed on the storage capacity,
size of basins of attraction,
and the superretrieval states.

\section{Model}
Let us consider a model with $N$ units.
Each unit has a binary variable $s_i\in\{-1,\,1\}$, $i=1,\,\ldots,\,N$,
and $\ss=[s_1,\,\ldots,\,s_N]\in\{-1,\,1\}^N$ represents a microscopic state
of the model.
Each unit stochastically and asynchronously updates
the value of $s_i$ based on the current value of the ``local field''
\begin{equation}
h_i(\ss)=\sum_{j\not=i}J_{ij}s_j,
\end{equation}
where $J_{ij}$ is a synaptic weight from neuron $j$ to neuron $i$.
The probability of state flip $s_i:=-s_i$ is given by
the following transition probability
\begin{equation}
w_i(\ss)={1\over2}(1-s_if(h_i(\ss))),
\end{equation}
where $f: \RR\mapsto[-1,\,1]$ is the output function.
\begin{figure}
\begin{center}
\unitlength=1pt
\begin{picture}(210,110)
\put(0,50){\vector(1,0){200}}
\put(100,0){\vector(0,1){100}}
\thicklines
\put(10,80){\line(1,0){45}}
\put(55,80){\line(0,-1){60}}
\put(55,20){\line(1,0){45}}
\put(100,20){\line(0,1){60}}
\put(100,80){\line(1,0){45}}
\put(145,80){\line(0,-1){60}}
\put(145,20){\line(1,0){45}}
\thinlines
\put(92,77){\makebox(0,0)[b]{1}}
\put(108,38){\makebox(0,0)[b]{0}}
\put(110,17){\makebox(0,0)[b]{$-1$}}
\put(44,54){\makebox(0,0)[b]{$-\theta$}}
\put(153,38){\makebox(0,0)[b]{$\theta$}}
\put(100,102){\makebox(0,0)[b]{$f(x)$}}
\put(205,47){\makebox(0,0)[b]{$x$}}
\end{picture}
\caption{Non-monotonic function $f(x)$.}
\label{fig:nonmono}
\end{center}
\end{figure}
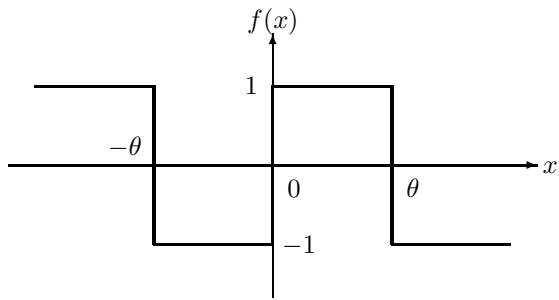
Taking $f(x)=\tanh\beta x$ yields the conventional Hopfield model.
In this paper we consider the following non-monotonic function
(Fig.~\ref{fig:nonmono}):
\begin{equation}
f(x)=\left\{\begin{array}{ll}
-\sgn(x) & (|x|\ge\theta)\\
 \sgn(x) & (|x|<\theta)
\end{array}\right.
\label{eq:nonmono}
\end{equation}
Its functional form is the same as that treated by Nishimori
and Opri\c{s}\cite{NOp93}.
Throughout the paper, we follow the common convention
about the time scale, that it is taken in such a way that
the average number per unit time (frequency) of updates each unit
executes is 1.

The model memorizes $p=\alpha N$ binary patterns
$\xxi^\mu=[\xi_1^\mu,\,\ldots,\,\xi_N^\mu]\in\{-1,\,1\}^N$,
$\mu=1,\,\ldots,\,p$,
via the Hebb rule,
\begin{equation}
J_{ij}={1\over N}\sum_{\mu=1}^p\xi_i^\mu\xi_j^\mu\quad(i\not=j).
\end{equation}
The quantity $\alpha\equiv p/N$ is called the memory rate.
We consider the case where the patterns to be memorized
are randomly generated, that is,
each of $\xi_i^\mu$'s takes the value $\pm1$ with probability $1/2$,
independently of others.

For measuring how well the model retrieves a pattern $\mu$,
the correlation, or overlap,
\begin{equation}
m^\mu(\ss)={1\over N}\sum_{i=1}^N\xi_i^\mu s_i
\end{equation}
is used:
$|m^\mu(\ss)|\le1$ holds by definition, and if $m^\mu(\ss)=1$, then
the model is in the state $\ss=\xxi^\mu$,
and it is regarded as perfectly retrieving the pattern $\mu$.
When $m^\mu(\ss)=-1$ the model is in the state $\ss=-\xxi^\mu$.
The state is called the reversal state, but it can also be regarded
as retrieving the pattern $\mu$ due to the symmetry of the model.

We assume that the model is going to retrieve one single pattern,
so that $m^\mu(\ss)$ are of order unity for that pattern only
(the {\em condensed ansatz}),
and that pattern $\mu=1$ is nominated for retrieval,
without loss of generality.
Then $m\equiv m^1(\ss)$ can be taken as a macroscopic variable,
or order parameter,
describing how well the model retrieves the nominated pattern.
If the model reaches an equilibrium with $m\not=0$,
the model is said to successfully retrieve the pattern,
and such an equilibrium is called a retrieval state.
The local field $h_i(\ss)$ is now rewritten as
\begin{equation}
h_i(\ss)=\xi_i^1[m+z_i(\ss)]-{1\over N}s_i,\quad
z_i(\ss)\equiv\xi_i^1\sum_{\mu>1}^p\xi_i^\mu
{1\over N}\sum_{j\not=i}\xi_k^\mu s_j.
\end{equation}
The term $z_i(\ss)$ represents the interference in the local field
from non-nominated patterns $\mu>1$, and is called the ``noise'' term.
Although time evolution of the microscopic state of the model
is stochastic in nature,
it is observed that the evolution of the order parameter $m$
in the course of pattern retrieval
can be often seen as being governed by a certain deterministic law.
To describe this retrieval dynamics is one of important problems
in this field.
Path integral formalism\cite{RSZ88} provides an exact description
to the retrieval dynamics,
but it requires parametrization with infinite degrees of freedom
and hence practically intractable.

\section{Two-parameter dynamical replica theory (2-DRT)}
2-DRT, proposed by Coolen and Sherrington\cite{cs-l,cs},
provides a tractable, and yet reasonably well for the conventional model,
description of the retrieval dynamics.
It uses two parameters, $m$ and $r$, as the order parameters,
the latter being defined as
\begin{equation}
r\equiv{1\over\alpha}\sum_{\mu>1}(m^\mu(\ss))^2,
\end{equation}
which intuitively represents the degree of interferential effect
of non-nominated patterns $\mu>1$ onto the retrieval of pattern 1.
2-DRT derives deterministic flow equations for these two
order parameters.
Without any assumptions, one cannot expect that the flow equations
for these two parameters are closed, and thus the time evolution
of $m$ and $r$ cannot be completely determined by current values of them.
In the framework of DRT in general, the following two assumptions are made:
\begin{enumerate}
\item Self-averaging of flow equations with respect to
	randomness of the system (randomly chosen patterns for the case
	treated in this paper).
\item Probability equipartitioning within subshells:
	when values of the order parameters are given,
	the probability distribution of the corresponding microscopic state
	can be regarded as being uniform over the subshell
	(the set of microscopic states
	which have the specified values of the order parameters),
	with regard to calculation of the flow equations.
\end{enumerate}

Owing to these closing assumptions, one can derive
the deterministic flow equations for $m$ and $r$:
\begin{eqnarray}
&&{dm\over dt}=\int dz\;D_{m,\,r}[z]\;f(m+z)-m \nonumber\\
&&{1\over2}{dr\over dt}
={1\over\alpha}\int dz\;D_{m,\,r}[z]\;zf(m+z)+1-r \nonumber
\end{eqnarray}
where $D_{m,\,r}[z]$ is the distribution of the noise terms.
Replica calculation gives, within the replica-symmetric (RS) ansatz,
the RS solution $D_{m,\,r}^{\rm RS}[z]$ for the noise distribution,
which has been derived by CS\cite{cs-l,cs} as
\begin{eqnarray}
D_{m,\,r}^{\rm RS}[z]&{}={}&
{e^{-(\Delta+z)^2/2\alpha r}\over2\sqrt{2\pi\alpha r}}
  \biggl\{1-\int Dy\tanh\biggl[
  \lambda y\left(\Delta\over\rho\alpha r\right)^{1/2}
  \!\!\!\!{}+(\Delta+z)\rho{r_\AGS\over r}+\mu\biggr]\biggr\}
  \nonumber\\
&&
{}+{e^{-(\Delta-z)^2/2\alpha r}\over2\sqrt{2\pi\alpha r}}
  \biggl\{1-\int Dy\tanh\biggl[
  \lambda y\left(\Delta\over\rho\alpha r\right)^{1/2}
  \!\!\!\!{}+(\Delta-z)\rho{r_\AGS\over r}-\mu\biggr]\biggr\}
\label{eq:CSnoisedist}
\end{eqnarray}
\begin{equation}
\Delta=\rho\alpha(r-r_\AGS)
\end{equation}
\begin{equation}
r_\AGS={\lambda^2\over\rho^2\alpha}
\end{equation}
where $Dy=(dy/\sqrt{2\pi})e^{-y^2/2}$ is the Gaussian measure.
The parameters $\{q,\,\lambda,\,\rho,\,\mu\}$ are
to be determined from $m$, $r$, and $\alpha$ by the following
saddle-point equations:
\begin{equation}
\begin{array}{ll}
\displaystyle r={1-\rho(1-q)^2\over[1-\rho(1-q)]^2},&
\displaystyle \lambda={\rho\sqrt{\alpha q}\over1-\rho(1-q)}\\
\\
\displaystyle m=\int Dy\,\tanh(\lambda y+\mu),&
\displaystyle q=\int Dy\,\tanh^2(\lambda y+\mu)
\end{array}
\label{eq:CS-speq}
\end{equation}

It should be noted that the replica calculation of the noise
distribution bears no relation to the dynamics of the model
and the choice of the output function $f$;
it executes averaging over the ($m$, $r$)-subshell
with uniform measure, and therefore
the calculation is not dynamical but configurational.
This is the reason why, apart from validity of the two assumptions,
2-DRT can be straightforwardly applied to the non-monotonic model.

The latter of the two above-mentioned assumptions,
the {\em equipartitioning\/} assumption,
is a critical one
because it has been known that
it is not valid for the conventional model\cite{Tanaka98},
as well as the continuous-valued linear system,
or the Langevin spin system\cite{CF94}.
To show this directly for the conventional model,
Tanaka and Osawa\cite{Tanaka98} have proposed a dynamics,
called ($m$, $r$)-annealing.
The ($m$, $r$)-annealing is defined,
on the basis of the solution $\{\rho,\,\mu\}$ of the saddle-point equations
for given $m$ and $r$,
as the dynamics of the conventional model
with the inverse temperature $\rho$ (that is,
it uses $f(x)=\tanh\rho x$), but an extra bias is added to the local field,
\begin{equation}
h_i(\ss)=\sum_{j\not=i}J_{ij}s_j+b\xi_i^1,
\end{equation}
where $b\equiv\mu/\rho-m$.
It has been shown that the ($m$, $r$)-annealing executes
Monte Carlo sampling from a ($m$, $r$)-subshell
with uniform probability (in the limit $N\to\infty$),
hence realizing the equipartitioning.
It is useful in investigating validity of the equipartitioning
assumption, and will be utilized in this paper.

\section{Results}
\subsection{Time evolution of order parameters}
\begin{figure}
\begin{center}
\begin{minipage}{60mm}
\begin{center}
\leavevmode
\epsfxsize=60mm \epsfbox{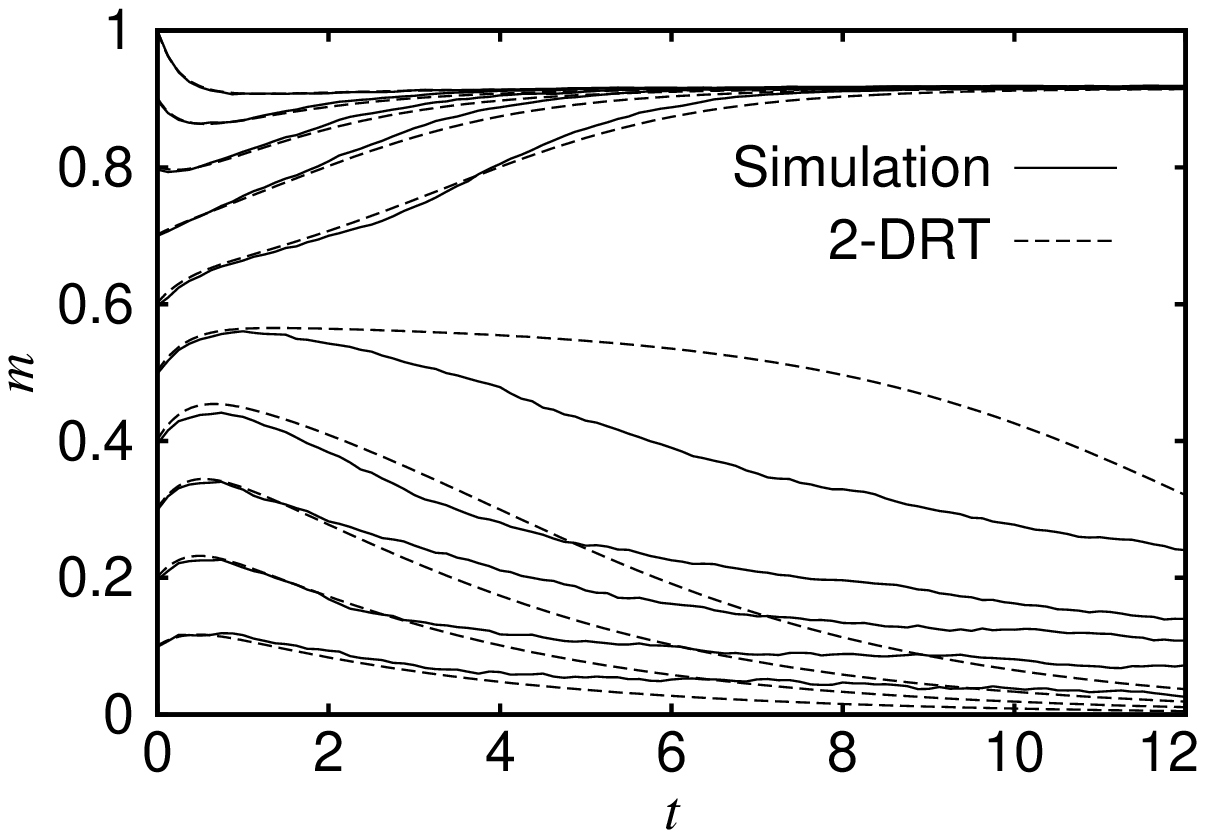}\\
(a) $m$-$t$ plot
\end{center}
\end{minipage}
\ %
\begin{minipage}{60mm}
\begin{center}
\leavevmode
\epsfxsize=60mm \epsfbox{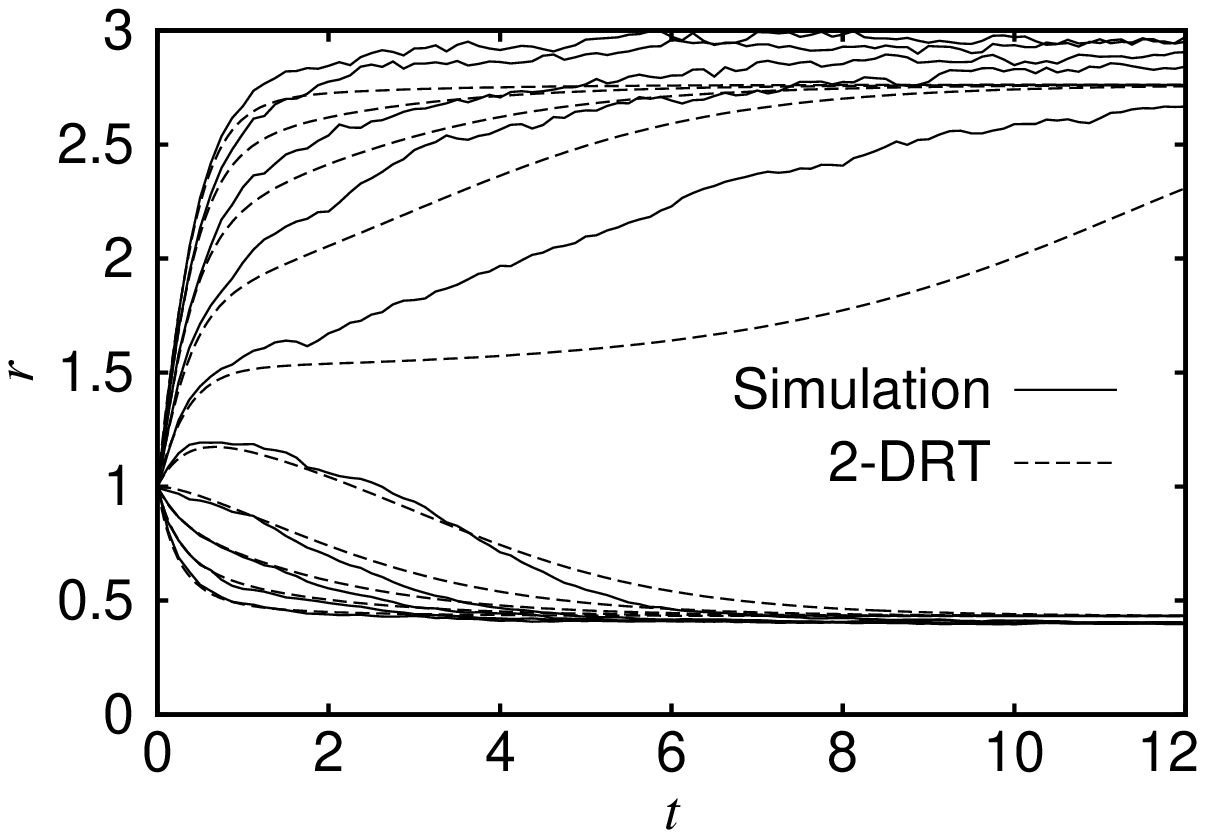}\\
(b) $r$-$t$ plot
\end{center}
\end{minipage}
\par\vspace{3mm}

\begin{minipage}{60mm}
\begin{center}
\leavevmode
\epsfxsize=60mm \epsfbox{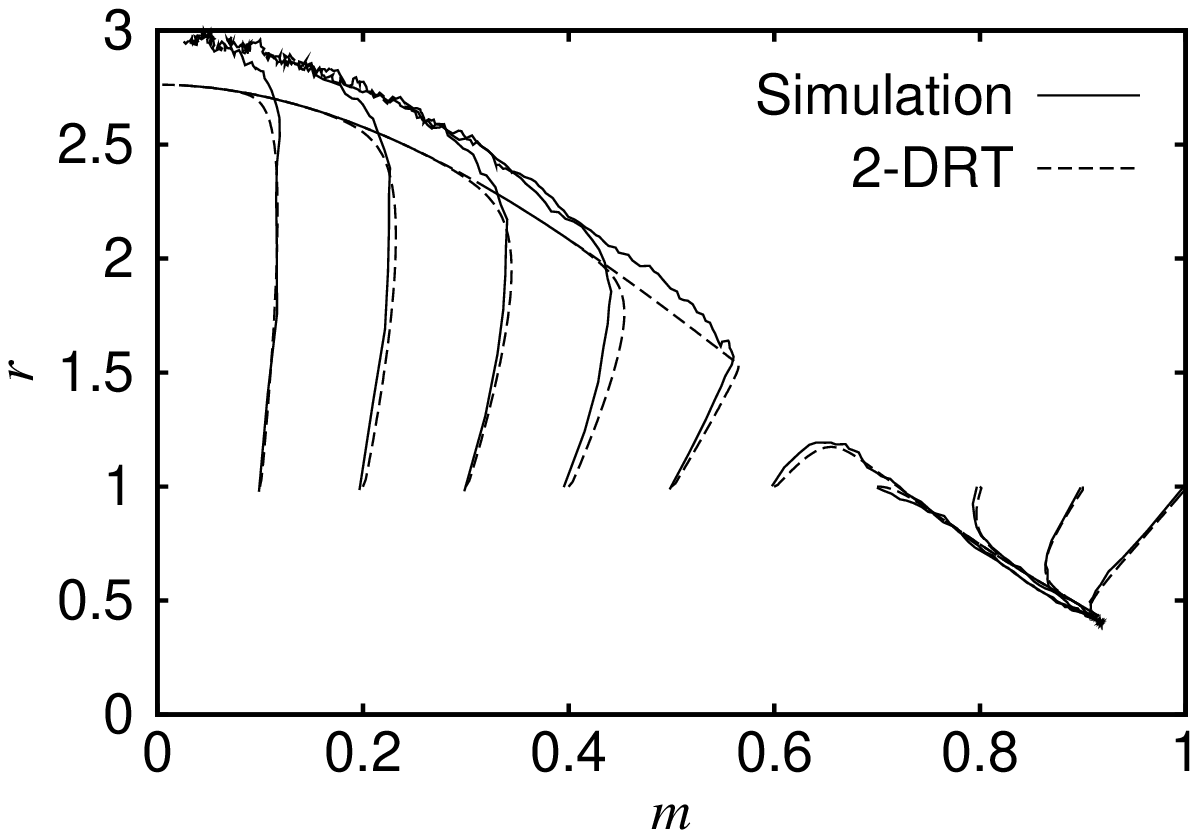}\\
(c) $m$-$r$ plot
\end{center}
\end{minipage}
\caption{Dynamics obtained by simulations with $N=2^{15}$ (solid)
	and computed by 2-DRT (dashed) for the case with $\alpha=0.2$
	and $\theta=1.4$.}
\label{fig:Dyn}
\end{center}
\end{figure}
\begin{figure}
\begin{center}
\epsfxsize=60mm \epsfbox{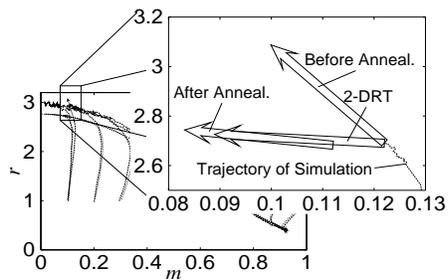}
\caption{Flow vectors $(\dot m,\,\dot r)$ at $t=1$
	on the trajectory starting at $(m,\,r)=(0.1,\,1)$
	for the model with $\alpha=0.2$, $\theta=1.4$ and $N=2^{15}$.
	Three flow vectors, ones before and after the ($m$, $r$)-annealing,
	and one computed by 2-DRT, are shown.}
\label{fig:FV}
\end{center}
\end{figure}
We first examined whether 2-DRT describes overall characteristics of
the dynamics of the non-monotonic model.
We assume the common convention
that initial microscopic states of the model are given by
randomly corrupting the nominated pattern; that is,
the initial state $\ss$ is set by the probability law
${\rm Prob}[s_i]=\delta(s_i-\xi_i^1)(1+m_0)/2
+\delta(s_i+\xi_i^1)(1-m_0)/2$, so that
the initial overlap $m(t=0)$ approximately equals to $m_0$
when $N$ is sufficiently large.
Following this initialization procedure, the initial value of $r$,
when $N$ is sufficiently large,
approximately equals to $1$.
We found that 2-DRT does describe the dynamics of the model considerably well,
as shown in Fig.~\ref{fig:Dyn} for the case with $\alpha=0.2$ and $\theta=1.4$.
2-DRT reproduced the trajectories almost exactly when retrieval succeeded.
Noticeable disagreement between simulations and 2-DRT was seen for cases
where retrieval failed.
Characteristic aspect of the disagreement is that
the trajectories predicted by 2-DRT exhibit, in early stages,
overall slowing down
against the corresponding simulations.
These observations are essentially the same as those found
in the conventional model\cite{cs,NOz93,Tanaka98}.
The observed disagreement is due to the failure of the equipartitioning
assumption of 2-DRT just as in the conventional model,
as demonstrated by the following numerical experiment.
The procedure of this experiment is as follows:
Simulate a model with $\theta=1.4$ and $\alpha=0.2$ by setting
its initial condition as $(m,\,r)=(0.1,\,1)$,
stop it at $t=1$,
and then execute the ($m$, $r$)-annealing for 80 units of time.
The flow vectors $(\dot m,\,\dot r)$, evaluated from
(1) the model just before executing the ($m$, $r$)-annealing,
(2) the model just after it, and
(3) 2-DRT,
were compared, and the result is summarized in Fig.~\ref{fig:FV}.
This shows that the flow vectors of the model after the ($m$, $r$)-annealing
and of 2-DRT are almost the same, whereas the flow vector before
the ($m$, $r$)-annealing is different from these two,
which means that the equipartitioning assumption does not hold in this case.

\subsection{Capacity and basins of attraction}
In this paper, we define the storage capacity $\alpha_c$ as the maximum value
of $\alpha$ for which a stable macroscopic state with nonvanishing $m$ exists.
We call an equilibrium macroscopic state with nonvanishing $m$
the retrieval state.
Since the retrieval state may be unstable,
the condition for the existence of a retrieval state
will give an overestimate of the true storage capacity.
On the other hand, a stable retrieval state may have a very small basin
of attraction, so that we may fail to find such a retrieval state
in numerical experiments even though it is stable.

\begin{figure}
\begin{center}
\epsfxsize=60mm \epsfbox{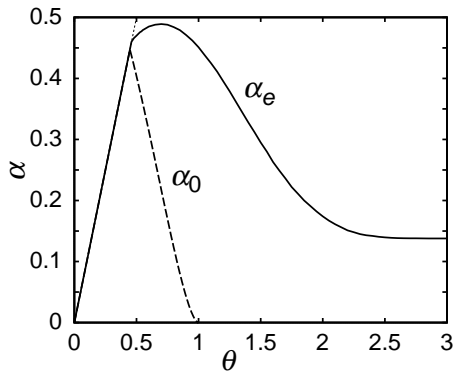}
\end{center}
\caption{Plot of $\alpha_e$, the maximum of $\alpha$ for which a retrieval
state exists (solid), and critical capacity
$\alpha_0$ (dashed), under which superretrieval occurs, against $\theta$,
evaluated by SCSNA calculation.
Thin dashed line shows $\alpha=\theta$, which limits
the capacity in the small-$\theta$ region.}
\label{fig:PhaseDiag}
\end{figure}
Shiino and Fukai\cite{SF93} have analyzed
equilibrium properties of
continuous-valued continuous-time non-monotonic models
using SCSNA.
Although we consider the model with binary variables in this paper
rather than ones with continuous values,
the equilibrium condition is shown to be the same as that
for the continuous-value continuous-time model
owing to the current choice of the output function $f$
(eq.~(\ref{eq:nonmono})),
and thus SCSNA can be applied to the model treated here.
We executed the SCSNA calculation on this model, and
Fig.~\ref{fig:PhaseDiag} shows the result
for $\alpha_e$, the maximum of $\alpha$ for which a retrieval state exists,
versus $\theta$.
When $\theta\to\infty$,
$\alpha_e$ approaches the well-known value 0.138,
confirming that SCSNA is consistent
with the Amit-Gutfreund-Sompolinsky (AGS) theory\cite{ags}.
As $\theta$ becomes smaller, $\alpha_e$ increases
so that it reaches its maximal value $\alpha_e=0.489$ at $\theta\approx0.7$.

As we have already discussed, $\alpha_e$ gives an overestimation
of the true storage capacity $\alpha_c$\cite{SF93}.
To evaluate $\alpha_c$ itself,
we have to take into account the dynamical aspect
of the retrieval process.
This discussion leads us to the idea to apply 2-DRT
to determine the storage capacity $\alpha_c$,
by observing whether or not the trajectories from arbitrary initial conditions
approach a retrieval state.

Two points have to be mentioned here.
First, although 2-DRT becomes exact at equilibrium
for the conventional model\cite{cs-l,cs}, it is no longer so
for the non-monotonic model,
which means that 2-DRT may not reproduce the storage capacity.
Second, since we assume the initialization procedure described above,
only the retrieval states which can be reachable from initial conditions
with $r=1$ are to be observed in the simulations.
To make correspondence with this experimental setup,
we estimated the storage capacity $\alpha_c$ by tracking 2-DRT trajectories
from initial conditions with $r=1$.
The storage capacity estimated by the simulations
and by the 2-DRT trajectory tracking may be therefore an underestimate
against the {\em true} storage capacity,
because there might be stable retrieval states unreachable from
any initial condition with $r=1$ (see Sect.~\ref{sec:sr}).

\begin{figure}
\begin{center}
\epsfxsize=60mm \epsfbox{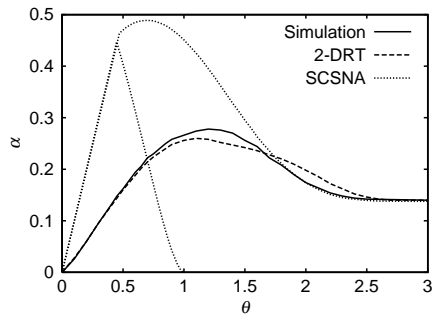}
\caption{Plot of storage capacity $\alpha_c$ against $\theta$,
evaluated by simulation (solid) and by 2-DRT trajectory tracking (dashed).
Plots of $\alpha_e$ and $\alpha_0$ evaluated by SCSNA
(shown in Figure \protect\ref{fig:PhaseDiag}) are also shown for comparison.}
\label{fig:PD-DRT}
\end{center}
\end{figure}
Figure~\ref{fig:PD-DRT} shows the estimated storage capacity $\alpha_c$
by 2-DRT trajectory tracking and by simulations.
It can be seen that 2-DRT trajectory tracking well reproduces
the storage capacity
obtained by the simulations for the whole range of $\theta$.
Comparing it with Fig.~\ref{fig:PhaseDiag} reveals that,
while the agreement is good when $\theta$ is large,
the discrepancy becomes apparent as $\theta$ becomes less than about $1.5$,
showing that the storage capacity estimated by 2-DRT is
considerably smaller than that estimated by SCSNA.
It may be partly explained by the overestimation of SCSNA described above.

\begin{figure}
\begin{center}
\begin{minipage}{60mm}
\begin{center}
\leavevmode
\epsfxsize=60mm \epsfbox{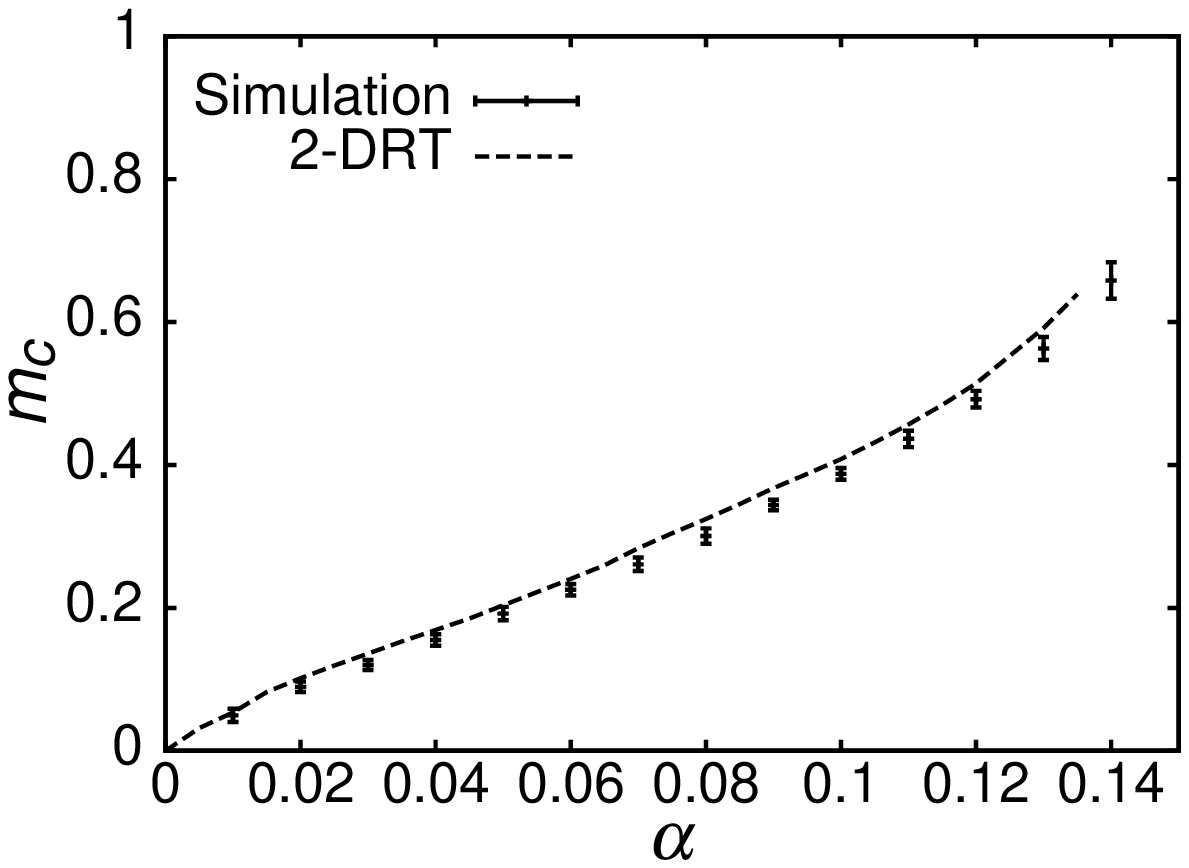}\\
(a) $\theta=+\infty$
\end{center}
\end{minipage}
\begin{minipage}{60mm}
\begin{center}
\leavevmode
\epsfxsize=60mm \epsfbox{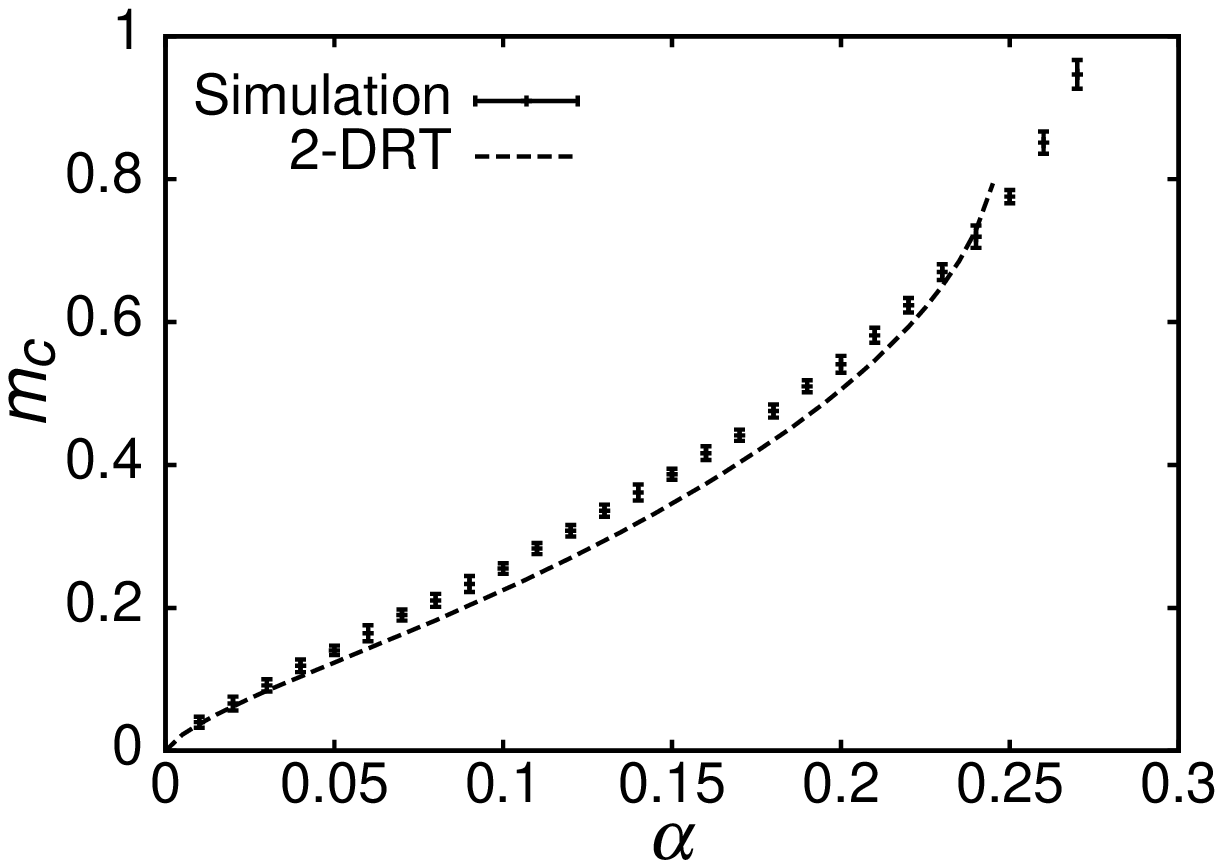}\\
(b) $\theta=1.4$
\end{center}
\end{minipage}
\begin{minipage}{60mm}
\begin{center}
\leavevmode
\epsfxsize=60mm \epsfbox{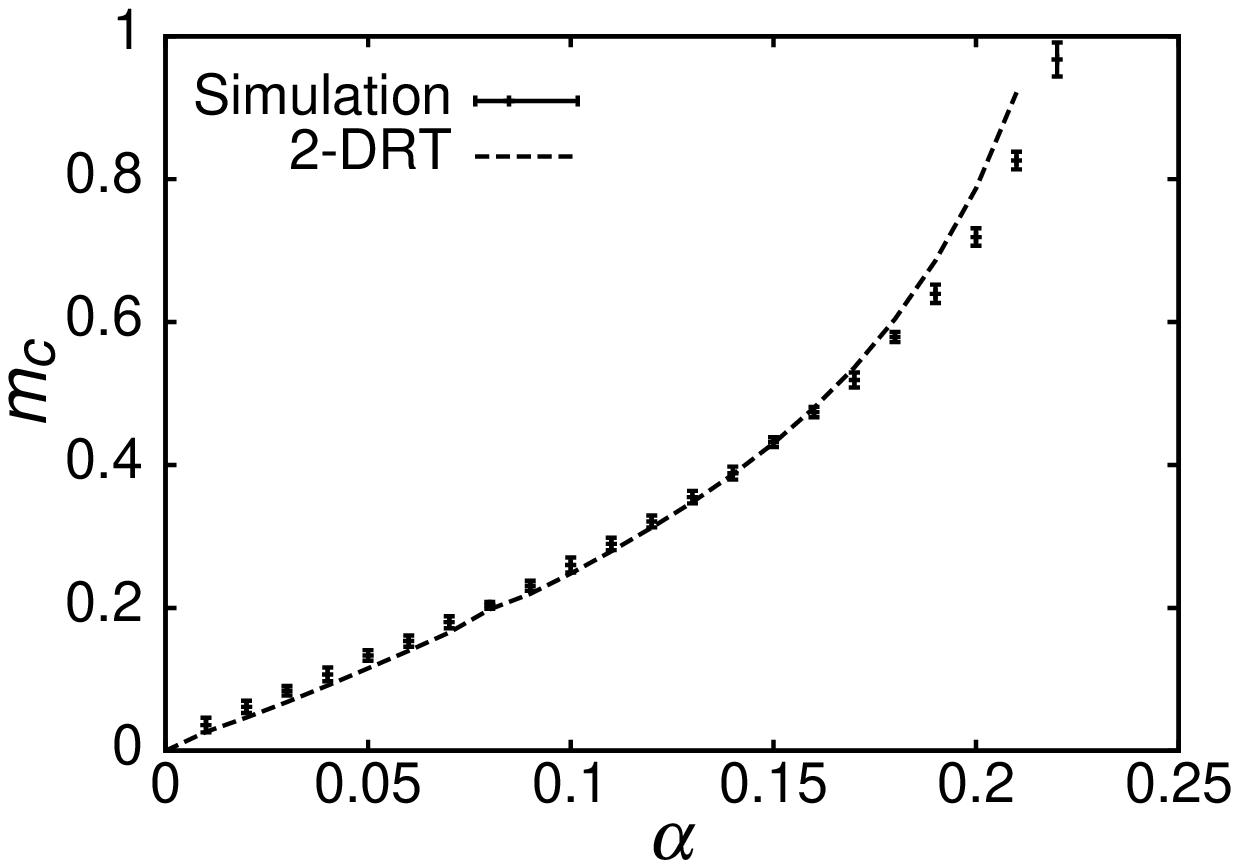}\\
(c) $\theta=0.7$
\end{center}
\end{minipage}
\caption{Plot of critical initial overlap $m_c$ against $\alpha$,
evaluated by 2-DRT (solid line) and by simulation (markers)
for $\theta=+\infty$, $1.4$ and $0.7$.}
\label{fig:basin}
\end{center}
\end{figure}

\begin{figure}
\begin{center}
\begin{minipage}{60mm}
\begin{center}
\leavevmode
\epsfxsize=60mm \epsfbox{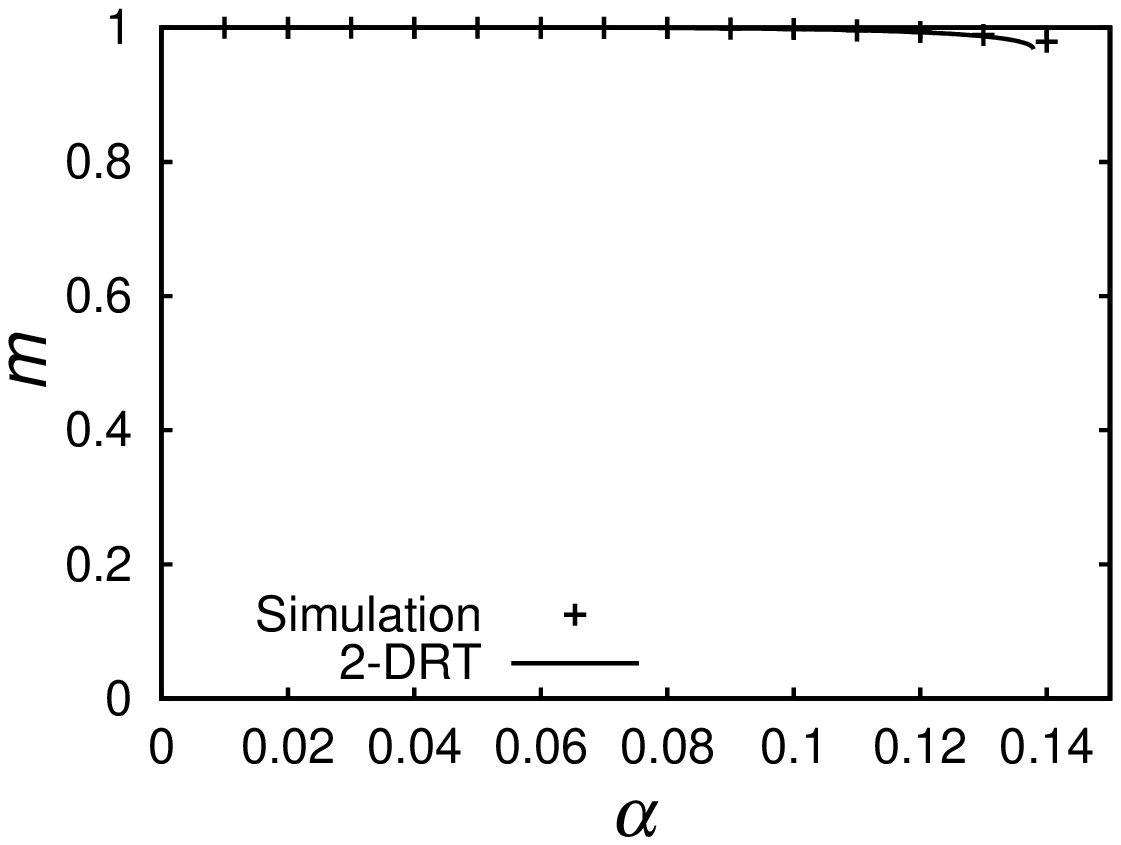}\\
(a) $\theta=+\infty$
\end{center}
\end{minipage}
\begin{minipage}{60mm}
\begin{center}
\leavevmode
\epsfxsize=60mm \epsfbox{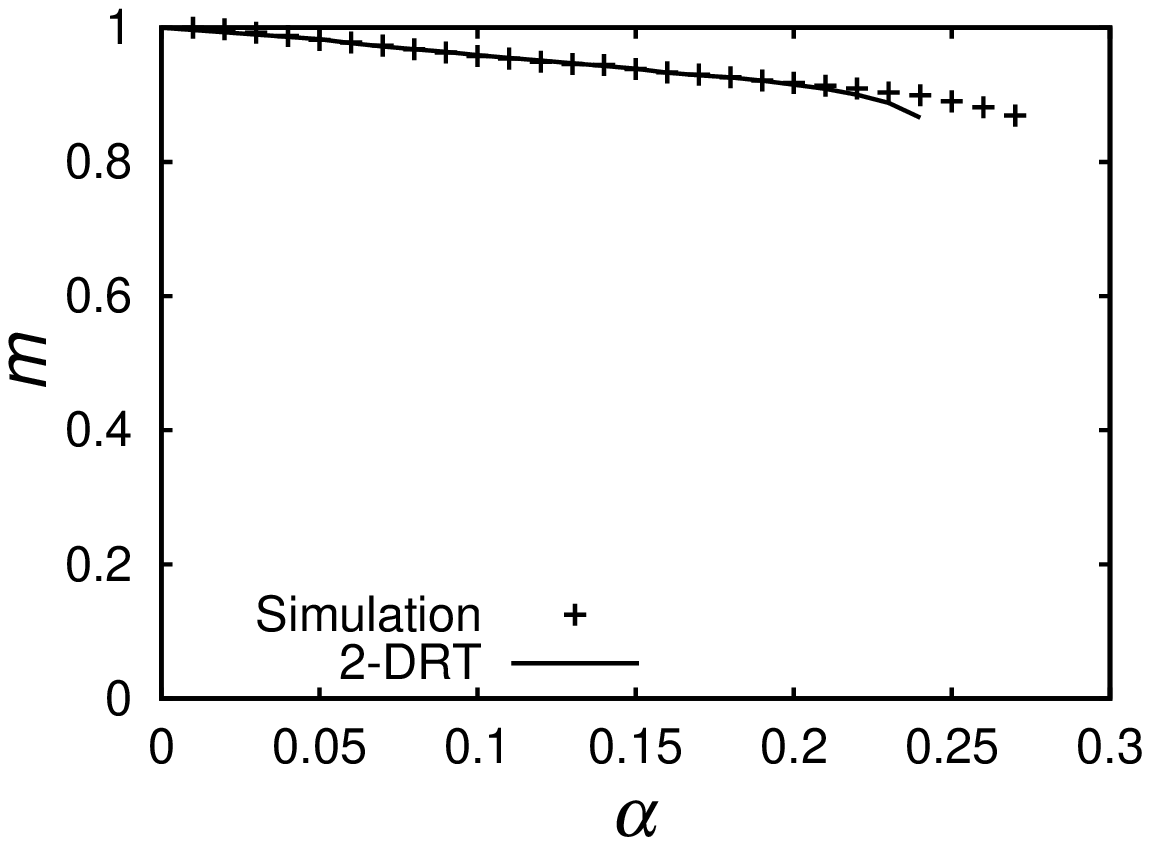}\\
(b) $\theta=1.4$
\end{center}
\end{minipage}
\begin{minipage}{60mm}
\begin{center}
\leavevmode
\epsfxsize=60mm \epsfbox{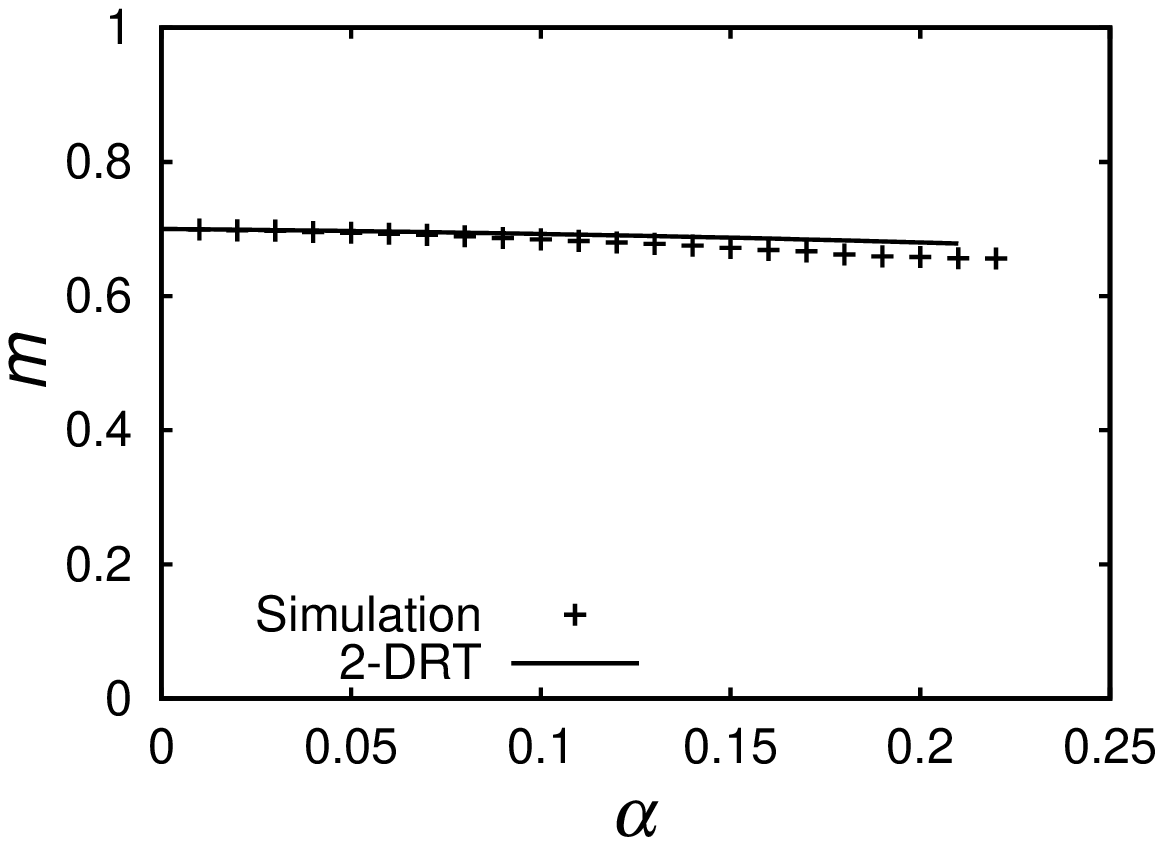}\\
(c) $\theta=0.7$
\end{center}
\end{minipage}
\caption{Plot of the value of $m$ of the retrieval state against $\alpha$,
evaluated by 2-DRT (solid line) and by simulation (markers)
for $\theta=+\infty$, $1.4$ and $0.7$.}
\label{fig:retrieval-m}
\end{center}
\end{figure}
One of the main advantage of dynamical theory is
that it allows us to evaluate basins of attraction,
because they are essentially of dynamical nature.
We used 2-DRT to evaluate the basin of attraction of the retrieval states.
Since we adopt the above-mentioned initialization convention,
we consider the basin of attraction as being represented
in terms of $m$ only, that is,
we regard a value of $m_0$ as belonging to the basin of attraction
of the retrieval state if the trajectory starting at the state
$(m,\,r)=(m_0,\,1)$ approaches the retrieval state.
Evaluating the basin of attraction defines the critical initial overlap $m_c$,
which means that initial states $(m_0,\,1)$ with $m_0\ge m_c$
yield successful retrieval.
Figures~\ref{fig:basin} and \ref{fig:retrieval-m} show
the critical initial overlap $m_c$
and the values of $m$ of the retrieval state, respectively,
evaluated by 2-DRT and by simulations.
It is readily seen that enlargement of the basin of attraction
occurs as $\theta$ becomes small,
and that 2-DRT captures this phenomenon reasonably well.

\subsection{Superretrieval states}
\label{sec:sr}
As a result of SCSNA analysis on non-monotonic models,
Shiino and Fukai\cite{SF93} have shown that there is
a phase where equilibrium states corresponding to ``perfect'' retrieval
exist.
Such states are called the superretrieval states,
whose existence is one of unique features of the non-monotonic models.
Here, ``perfect'' means that the correlation
of the sign of the local field $h_i(\ss)$
(not of $\ss$)
with the nominated pattern $\xxi^1$ is exactly equal to $\pm1$.
The correlation defined as above is called the tolerance overlap\cite{SF93}.
The critical capacity $\alpha_0$, below which the superretrieval occurs,
can be evaluated numerically by SCSNA, and is also shown in
Fig.~\ref{fig:PhaseDiag}.

An explanation, given by Shiino and Fukai\cite{SF93},
for possibility of such states is briefly as follows:
For such states $r_\AGS\to+0$ holds, which has been confirmed by
numerically solving relevant self-consistent equations.
Since variance of the noise term (without the ``systematic'' term,
$\Gamma Y$, in their terminology\cite{SF93}) is given by $\alpha r_\AGS$
in SCSNA, it implies that the effect of the noise completely vanishes
in the states, which enables the superretrieval.

\begin{figure}
\begin{center}
\epsfxsize=60mm \epsfbox{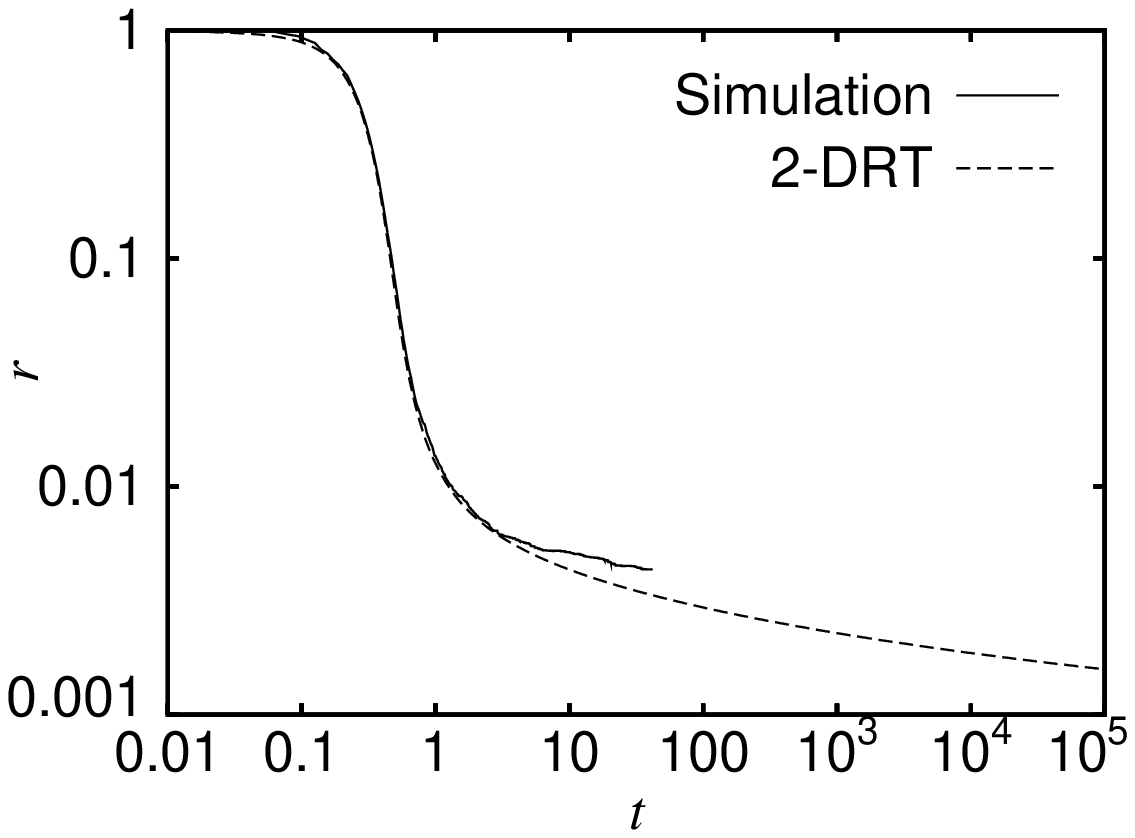}
\caption{Time evolution of $r$ for $\theta=0.4$, $\alpha=0.05$,
	and with initial condition $m_0=0.9$,
	evaluated by simulation with $N=2^{15}$ (solid) and by 2-DRT (dashed).}
\label{fig:SR-r-t}
\end{center}
\end{figure}
\begin{figure}
\begin{center}
\epsfxsize=60mm \epsfbox{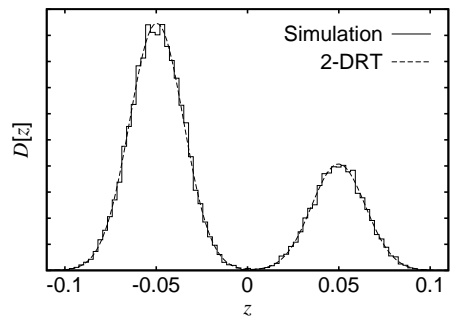}
\caption{Noise distribution $D[z]$ at the equilibrium state
	($(m,\,r)=(0.398,\,0.00440)$) achieved by
	the simulation with $\theta=0.4$, $\alpha=0.05$, $N=2^{15}$,
	and initial condition $m_0=0.9$ (solid),
	and the one computed by 2-DRT for the same values of
	$(m,\,r)$ (dashed).}
\label{fig:SR-noise-dist}
\end{center}
\end{figure}
\begin{figure}
\begin{center}
\epsfxsize=60mm \epsfbox{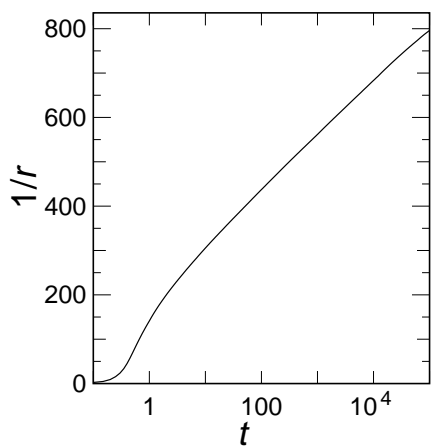}
\caption{Plot of $1/r$ versus $\ln t$ for $\theta=0.4$, $\alpha=0.05$,
	and with initial condition $m_0=0.9$,
	evaluated by 2-DRT.}
\label{fig:rt}
\end{center}
\end{figure}
Especially interesting is whether or not 2-DRT provides
an appropriate description of dynamics
which is bound for superretrieval states.
To see this, we examined the case where $\theta=0.4$ and $\alpha=0.05$,
which, according to SCSNA analysis,
is expected to have the superretrieval states.
Figure~\ref{fig:SR-r-t} shows time evolution of $r$
evaluated by simulation and by 2-DRT,
with initial condition $m_0=0.9$.
Again, 2-DRT reproduced the simulation result fairly well.
For the simulation, state transition ended (confirmed numerically)
at $t\approx30$, where $(m,\,r)=(0.398,\,0.00440)$.
The tolerance overlap was evaluated to be exactly equal to $1$ at this state,
indicating that this is the superretrieval state.
Figure~\ref{fig:SR-noise-dist} shows the noise distribution $D[z]$
at the equilibrium state achieved by the simulation,
and the one computed by 2-DRT for the same values of $(m,\,r)$.
They are in good agreement, suggesting that 2-DRT can successfully predict
the trajectories even in the case of superretrieval,
provided that the system size $N$ is sufficiently large.
As for 2-DRT, the value of $r$ continued to decrease at $t\approx10^5$
(we stopped computation of 2-DRT at $t=10^5$ because rounding errors became
profound beyond this point),
where $(m,\,r)=(0.399,\,0.00159)$.
We observed numerically that $t$-dependence of the decrease of $r$
can be expressed,
to a good approximation, as $r(t)\propto1/\ln t$ (Fig.~\ref{fig:rt}),
which strongly supports the conjecture
that the trajectory obtained by 2-DRT certainly approaches $r=0$.
As can be seen by eq.~(\ref{eq:CSnoisedist}),
variance of the noise term is, roughly speaking, given by $\alpha r$
in 2-DRT.
Then, if the superretrieval states are described appropriately by 2-DRT,
they should correspond to the states with $r=+0$.
We therefore investigate in the following
the solutions of the saddle-point equations (\ref{eq:CS-speq})
when $r=+0$.

By formally taking the limit $r\to+0$,
the saddle-point equations (\ref{eq:CS-speq})
are reduced to the following equations
\begin{eqnarray}
\rho&=&-\infty\\
m&=&\int Dy\,\tanh(\lambda y+\mu)\\
q&=&\int Dy\,\tanh^2(\lambda y+\mu)\\
\lambda&=&-{\sqrt{\alpha q}\over1-q},
\end{eqnarray}
and, correspondingly, the RS noise distribution $D_{m,\,r=+0}^{\rm RS}[z]$
(\ref{eq:CSnoisedist}) becomes
\begin{equation}
D_{m,\,r=+0}^{\rm RS}[z]={1-m\over2}\delta(z-\alpha)
+{1+m\over2}\delta(z+\alpha).
\end{equation}
The condition that the macroscopic state $(m,\,r=+0)$ is an equilibrium state,
that is, $(\dot m,\,\dot r)=(0,\,0)$ holds,
is thus given by
\begin{equation}
f(m\pm\alpha)=\mp1.
\end{equation}
For the current choice of the function $f$ (eq.~(\ref{eq:nonmono})),
this condition is satisfied when $0<m-\alpha<\theta<m+\alpha$,
or equivalently,
$\max\{\theta-\alpha,\,\alpha\}<m<\min\{\theta+\alpha,\,1\}$.
This is in consistent with the observation from
the simulations that $m$ tends to approach $\theta$ when the superretrieval
occurs.

The applicability of the argument above depends
not only on the two assumptions made at the beginning
but also on the two following points:
The first one concerns the so-called freezing line,
which defines the points in the $(m,\,r)$ plane
where the number of microscopic states within the $(m,\,r)$-subshell
changes from an exponentially large number to an exponentially small
number in terms of $N$.
The second one concerns the so-called de Almeida-Thouless (AT) line\cite{AT78},
at which the replica-symmetry-breaking (RSB) occurs
and thus the RS ansatz becomes no longer valid.

\begin{figure}
\begin{center}
\epsfxsize=60mm \epsfbox{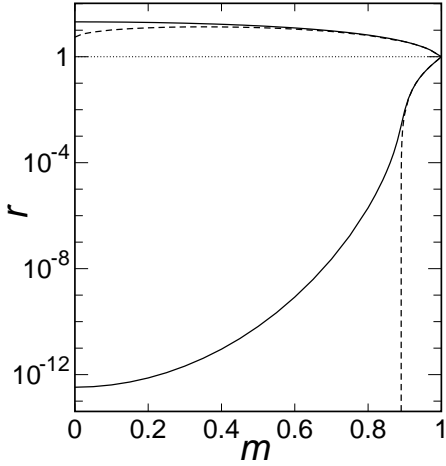}
\caption{The freezing line (solid) and the AT line (dashed) for $\alpha=0.05$.}
\label{fig:a=0.05}
\end{center}
\end{figure}
The freezing line is given, under the RS ansatz,
by the following equation\cite{cs}:
\begin{equation}
\int Dy\ln\cosh[\lambda y+\mu]-\mu m
-{\alpha\over2}\left[\ln[1-\rho(1-q)]
+{\rho(1-q)(1-\rho+3q\rho)\over[1-\rho(1-q)]^2}\right]+\ln2=0
\label{eq:freeze}
\end{equation}
The number of microscopic states within the subshell is exponentially large,
and therefore taking averages over the subshell is expected
to have the proper meaning,
as long as the left-hand side of eq.~(\ref{eq:freeze}) is positive.
For $r=1$ and $|m|<1$ the left-hand side becomes
\begin{equation}
-{1\over2}\left[(1+m)\ln{1+m\over2}+(1-m)\ln{1-m\over2}\right]>0,
\end{equation}
so that the number of microscopic states is indeed exponentially large.
In the limit $r\to+0$, however, the left-hand side of eq.~(\ref{eq:freeze})
goes to $-\infty$, implying that $r=0$ is outside the freezing line.
This means that there are so few microscopic states near $r=0$,
which may in part explain the observation from the simulations
that the trajectories approaching $r=0$
reach equilibrium before actually arriving at $r=0$.
That $r=0$ is outside the freezing line also means
that the argument with the formal limit $r\to+0$
eventually loses its proper meaning,
because the relevant subshell average is over an {\em exponentially small\/}
number of microscopic states.
Numerical evaluation reveals, however,
that the freezing line for $r<1$ lies
very close to $r=0$ (Fig.~\ref{fig:a=0.05}).
Moreover, the saddle-point solution of the order parameters
changes smoothly as $r$ tends to $+0$:
Let $\mu_0$, $q_0$ be the asymptotic values of the saddle-point solution
$\mu$, $q$, respectively.
Then the asymptotic form of the saddle-point solution as $r\to+0$ is given by
\begin{eqnarray}
\rho&=&-r^{-1}+O(1)\nonumber\\
\lambda&=&-{\sqrt{\alpha q_0}\over1-q_0}+O(r)\nonumber\\
\mu&=&\mu_0+O(r)\nonumber\\
q&=&q_0+O(r)\nonumber\\
\Delta&=&-\alpha+O(r).
\label{eq:asympt-sp}
\end{eqnarray}
We can therefore expect that the argument presented above
on the formal limit $r\to+0$ well captures qualitative aspects
of the equilibrium states achieved by the simulations.

The AT line is determined by examining stability of the RS solution.
Assuming that RSB is caused by destabilization of
the so-called ``replicon'' modes\cite{AGS87,FH91,AT78} for the case $r<1$
just as it has been assumed for the case $r\ge1$\cite{cs},
the AT line turns out to be given by the same formula
as given in \cite{cs} for $r\ge1$:
\begin{equation}
\alpha=\rho^2(\alpha+\Delta)^2\int{Dy\over\cosh^4(\lambda y+\mu)}
\label{eq:ATline}
\end{equation}
The RS solution is valid if the right-hand side of eq.~(\ref{eq:ATline})
is less than $\alpha$.
We confirmed that, for the case where $\alpha=0.05$,
the AT line lies in the region $m>0.890$ for $r<1$ (Fig.~\ref{fig:a=0.05}),
which implies that the superretrieval observed in the simulations
was irrelevant to RSB.

\begin{figure}
\begin{center}
\epsfxsize=60mm \epsfbox{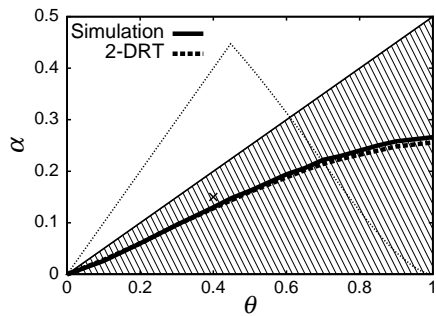}
\caption{The region where
the superretrieval solution $(m,\,r)=(\theta,\,0)$
is stable, evaluated by 2-DRT local stability analysis 
(shaded region), by tracking simulation trajectories
(region below thick solid curve) and by tracking 2-DRT trajectories 
(region below thick dashed curve).
The region where superretrieval state exists, evaluated by SCSNA
(see Fig.~\protect\ref{fig:PhaseDiag}) is also shown (dotted) for comparison.}
\label{fig:regs}
\end{center}
\end{figure}
We have conducted local stability analysis of the 2-DRT stationary
solutions $r=0$, $\max\{\theta-\alpha,\,\alpha\}<m<\min\{\theta+\alpha,\,1\}$
corresponding to the superretrieval state.
The result of the analysis states that,
among the superretrieval solutions,
$(m,\,r)=(\theta,\,0)$ is the only attractor when $2\alpha<\theta<1$,
although it becomes unstable when $\alpha<\theta<\min\{2\alpha,\,1\}$.
For the details of the local stability analysis, see Appendix.
In the simulations, however, it is not
for all $(\theta,\,\alpha)$ values satisfying $2\alpha<\theta<1$
that the superretrieval was observed.
Figure~\ref{fig:regs} shows the region where the local stability analysis
of the 2-DRT predicts the stable superretrieval state,
the one where the superretrieval is observed by simulations
(in the sense that tolerance overlap is numerically evaluated to be 1),
and the one where the superretrieval is predicted by tracking
the 2-DRT trajectories starting at $(m,\,r)=(m_0,\,1)$.
The result of 2-DRT trajectory tracking and that of simulations
are in good agreement with each other.
Both are within the region where the superretrieval state is
locally stable, as they should be, but apparently they do not coincide.
The region where the superretrieval is observed in simulations
may be further restricted by the following factors:
\begin{itemize}
\item The superretrieval solution $(m,\,r)=(\theta,\,0)$ may not
be reachable from the conventional initial states with $r=1$,
even though it is an attractor.
\item The superretrieval solution $(m,\,r)=(\theta,\,0)$ may be
at the outside of the RS region, where the stationarity
and local stability arguments, both based on the RS ansatz,
are no longer valid.
\end{itemize}

\begin{figure}
\begin{center}
\epsfxsize=60mm \epsfbox{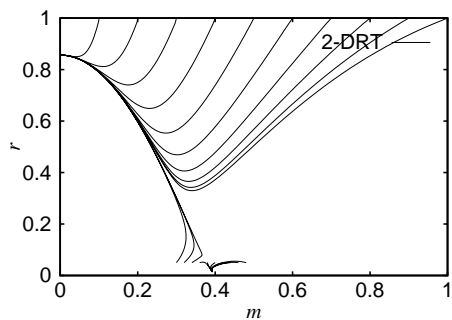}
\caption{Trajectories predicted by 2-DRT in $(m,\,r)$ plane
with $(\theta,\,\alpha)=(0.4,\,0.15)$.}
\label{fig:sample-a015h04}
\end{center}
\end{figure}
A demonstration regarding the former factor is shown in
Fig.~\ref{fig:sample-a015h04}.
For the condition $(\theta,\,\alpha)=(0.4,\,0.15)$
(marked by a cross in Fig.~\ref{fig:regs}), for example,
the superretrieval state is not observed
by following time evolution by either numerical simulation or 2-DRT.
Nevertheless, 2-DRT predicts that under this condition
the stable superretrieval state exists at $(m,\,r)=(0.4,\,0)$.
As shown in the figure,
2-DRT trajectory tracking shows that in this condition
the superretrieval state is indeed stable, but it is not reachable
from the initial states with $r=1$.

\begin{figure}
\begin{center}
\epsfxsize=60mm \epsfbox{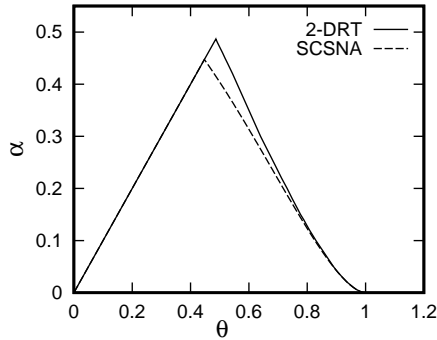}
\caption{The region where, according to 2-DRT,
the superretrieval solution $(m,\,r)=(\theta,\,0)$
is a stationary point and satisfies the RS ansatz (solid)
and the superretrieval phase evaluated by SCSNA (dashed).}
\label{fig:sr-cs-scsna}
\end{center}
\end{figure}

An interesting observation related to the latter factor is that
there is a rough numerical correspondence between
the region where the equilibrium superretrieval solution
$(m,\,r)=(\theta,\,0)$, $\alpha<\theta<1$, which may not be stable,
satisfies the RS ansatz,
and the region where SCSNA predicts superretrieval to occur,
as shown in Fig.~\ref{fig:sr-cs-scsna}.

\section{Conclusion}
We have studied the question of how well 2-DRT describes retrieval dynamics
of the non-monotonic model.
Although there is no theoretical justification for 2-DRT to be exact
either for the non-monotonic model,
2-DRT turns out to reproduce the retrieval dynamics quite well,
and it gives reasonable results as for the capacity,
basins of attractions, and the superretrieval states.


\appendix
\section{Local Stability Analysis of Superretrieval States}
We first split the RS noise distribution $D_{m,\,r}^{\rm RS}[z]$
into two components, as follows:
\begin{equation}
D_{m,\,r}^{\rm RS}[z]\equiv D^-[z]+D^+[z]
\end{equation}
\begin{equation}
D^\pm[z]={e^{-(\Delta\pm z)^2/2\alpha r}\over2\sqrt{2\pi\alpha r}}w_\pm(z)
\end{equation}
\begin{equation}
w_\pm(z)=1-\int Dy\,\tanh\biggl[
  \lambda y\left(\Delta\over\rho\alpha r\right)^{1/2}
  \!\!\!\!{}+(\Delta\pm z)\rho{r_\AGS\over r}\pm\mu\biggr]
\end{equation}
Note that $w^\pm(z)$ is ``slowly varying'' with respect to $z$, because
\begin{equation}
\left|dw^\pm(z)\over dz\right|<\left|\rho{r_\AGS\over r}\right|
\end{equation}
holds and the bound $|\rho r_\AGS/r|$ remains finite even when $r\to+0$.
We can thus regard that each component is basically a gaussian distribution
centered at $z=\mp\Delta$ and width $O(\sqrt{\alpha r})$,
and it has been modulated by a bounded, monotonic, and slowly varying function
$0<w_\pm(z)<2$.
From the asymptotic form of the saddle-point solution as $r\to+0$
(eq.~(\ref{eq:asympt-sp})),
we can expect, for small $r$, that the noise components $D^\pm[z]$ become
sharply peaked around $z=\mp\Delta\approx\pm\alpha$.
In the limit $r\to+0$, we have
\begin{equation}
D^\pm[z]={1\mp m\over2}\delta(z\mp\alpha),
\end{equation}
so that the condition
\begin{equation}
\max\{\theta-\alpha,\,\alpha\}<m<\min\{\theta+\alpha,\,1\}
\label{eq:cond-sr}
\end{equation}
is obtained for the existence of equilibrium states
of the form $(m,\,r)=(m,\,0)$, as discussed in Sect.~\ref{sec:sr}.

In this section we analyze local stability of the equilibrium states
$(m,\,r=0)$ satisfying the condition (\ref{eq:cond-sr}).
Using the noise components, the time evolution equations are rewritten as
\begin{eqnarray}
&&\dot m=\int_{-\infty}^\infty dz\,D^-[z]f(m+z)
+\int_{-\infty}^\infty dz\,D^+[z]f(m+z)-m\nonumber\\
&&{1\over2}\dot r={1\over\alpha}\left[
\int_{-\infty}^\infty dz\,D^-[z]\,zf(m+z)
+\int_{-\infty}^\infty dz\,D^+[z]\,zf(m+z)\right]+1-r.
\end{eqnarray}
Because $D^\pm[z]$ are sharply peaked,
as the first step of approximation we can assume that
$f(m+z)=f(m\mp\Delta)$ in the integrals with $D^\pm[z]$.
This assumption becomes exact in the limit $r\to+0$ and
when $f(m+z)$ is continuous around $z=\pm\alpha$,
but for finite $r$ it gives an approximate result
and the approximation error comes from the contribution
of the {\em tails} of $D^\pm[z]$ where $f(m+z)$ changes the sign.
For explanation purposes we introduce the following four regions:
\begin{eqnarray}
{\rm I}&\equiv&\{z\mid m+z<-\theta\}\nonumber\\
{\rm II}&\equiv&\{z\mid -\theta<m+z<0\}\nonumber\\
{\rm III}&\equiv&\{z\mid 0<m+z<\theta\}\nonumber\\
{\rm IV}&\equiv&\{z\mid \theta<m+z\}
\end{eqnarray}
$f(m+z)=1$ for $z\in{\rm I}$ or ${\rm III}$, and
$f(m+z)=-1$ for $z\in{\rm II}$ or ${\rm IV}$.
The equilibrium states which we are interested in
correspond to the case where the peak of $D^+[z]$ is in the region IV
and that of $D^-[z]$ in the region III.
In this case the time evolution equations are approximated to be
\begin{eqnarray}
&&\dot m\approx\int_{-\infty}^\infty dz\,D^-[z]
-\int_{-\infty}^\infty dz\,D^+[z]-m\nonumber\\
&&{1\over2}\dot r\approx{1\over\alpha}\left[
\int_{-\infty}^\infty dz\,D^-[z]\,z
-\int_{-\infty}^\infty dz\,D^+[z]\,z\right]+1-r.
\end{eqnarray}
However, direct calculation shows that the right-hand sides
of these equations exactly equal to 0.
This fact indicates that the time evolution near $r=0$ should be
governed by the contribution of the {\em tails}.

The principal contribution comes from the largest one of the following
three quantities:
\begin{enumerate}
\item Contribution of the tail of $D^+[z]$ in the region III:
\begin{equation}
I_1=2\int_{\rm III}dz\,D^+[z]\,a(z)
\approx 2\int_{-\infty}^{z_0}dz\,D^+[z]\,a(z)
\end{equation}
\item Contribution of the tail of $D^-[z]$ in the region IV:
\begin{equation}
I_2=-2\int_{\rm IV}dz\,D^-[z]\,a(z)
=-2\int_{z_0}^\infty dz\,D^-[z]\,a(z)
\end{equation}
\item Contribution of the tail of $D^-[z]$ in the region II:
\begin{equation}
I_3=-2\int_{\rm II}dz\,D^-[z]\,a(z)
\approx-2\int_{-\infty}^{-m} dz\,D^-[z]\,a(z)
\end{equation}
\end{enumerate}
where $z_0\equiv\theta-m$, and $a(z)=1$~or~$z$, depending on
which of $\dot m$ and $\dot r$ we are considering.
We approximate each contribution by extending
the integral region to $\infty$ or $-\infty$.
In fact this approximation does not affect the final result
in the $r\to+0$ limit
because it changes each quantity by a vanishingly small amount.

First let us consider the contribution to $\dot m$.
Evaluation for $I_1$ yields
\begin{eqnarray}
I_1&\approx&2\int_{-\infty}^{z_0}dz\,D^+[z]\nonumber\\
&=&\int_{-\infty}^{z_0}dz
{e^{-(\Delta+z)^2/2\alpha r}\over\sqrt{2\pi\alpha r}}w_+(z)\nonumber\\
&\approx&\int_{-\infty}^{z_0}dz
{e^{-(\Delta+z)^2/2\alpha r}\over\sqrt{2\pi\alpha r}}w_+(z_0)\nonumber\\
&\approx&{1\over\sqrt{2\pi}}\exp\left[
-{(z_0+\Delta)^2\over2\alpha r}-\ln{z_0+\Delta\over\sqrt{\alpha r}}
+\ln w_+(z_0)\right].
\end{eqnarray}
Similarly, for $I_2$ and $I_3$, we have
\begin{eqnarray}
I_2&\approx&-{1\over\sqrt{2\pi}}\exp\left[
-{(z_0-\Delta)^2\over2\alpha r}-\ln{z_0-\Delta\over\sqrt{\alpha r}}
+\ln w_-(z_0)\right],\\
I_3&\approx&-{1\over\sqrt{2\pi}}\exp\left[
-{(-m-\Delta)^2\over2\alpha r}-\ln{-m-\Delta\over\sqrt{\alpha r}}
+\ln w_-(-m)\right],
\end{eqnarray}
respectively.
In the $r\to+0$ limit, the dominant contribution comes from the first term
of the exponent for each case,
so that comparison of the term is sufficient to determine
which of $I_1$, $I_2$, and $I_3$ has the largest contribution to $\dot m$.
The result of the comparison for small $r$ is summarized as follows:
\begin{itemize}
\item When $-2\Delta<\theta$, the largest contribution comes from $I_1$ or
$I_2$.
If $m<\theta$, $I_1$ is the largest and $\dot m>0$.
Otherwise, $I_2$ is the largest and $\dot m<0$.
\item When $-2\Delta>\theta$, the largest contribution comes from $I_2$ or
$I_3$.
Since both $I_2$ and $I_3$ have negative contribution, $\dot m<0$.
\end{itemize}
From this result, we can conclude that the stable superretrieval state,
if it exists, should be $(m,\,r)=(\theta,\,0)$,
and that $2\alpha<\theta<1$ is a necessary condition
for the existence of the stable superretrieval state.

Let us now take a closer look at the flow near the state
$(m,\,r)=(\theta,\,0)$.
We let $\varepsilon\equiv-z_0=m-\theta$,
and consider time evolution of the two small quantities, $\varepsilon$ and $r$.
In the following arguments we assume that
$-2\Delta<\theta$ holds, so that the principal contribution
comes from $I_1$ or $I_2$, but not from $I_3$.
Under this assumption, we have
\begin{eqnarray}
\dot\varepsilon
&\approx&\sqrt{2\alpha r\over\pi}e^{-(\Delta^2+\varepsilon^2)/2\alpha r}
\bar w\Bigl(-{1\over\Delta}\Bigr)
\Bigl[\tanh{\Delta\varepsilon\over\alpha r}+{\delta\over\bar w}\Bigr]
\cosh{\Delta\varepsilon\over\alpha r}\nonumber\\
{1\over2}\dot r
&\approx&\sqrt{2\alpha r\over\pi}e^{-(\Delta^2+\varepsilon^2)/2\alpha r}
\bar w\Bigl({r\over\Delta^2}\Bigr)
\Bigl[\Bigl({\Delta\varepsilon\over\alpha r}-{\delta\over\bar w}\Bigr)
\tanh{\Delta\varepsilon\over\alpha r}
+\Bigl({\delta\over\bar w}{\Delta\varepsilon\over\alpha r}-1\Bigr)\Bigr]
\cosh{\Delta\varepsilon\over\alpha r},
\end{eqnarray}
where
\begin{eqnarray}
\bar w&\equiv&{1\over2}\bigl(w_+(z_0)+w_-(z_0)\bigr)\nonumber\\
\delta&\equiv&{1\over2}\bigl(w_+(z_0)-w_-(z_0)\bigr).
\end{eqnarray}
Note that $-1<\delta/\bar w<1$ holds.

Assuming that $\varepsilon/r$ remains finite,
we can readily see that
$\dot r$ is smaller in magnitude than $\dot\varepsilon$ by a factor $r$.
Then for small enough $r$ the slaving principle applies
and $\varepsilon$ is expected to relax toward its equilibrium value
much faster than $r$.
This justifies the adiabatic approximation, and we can regard that
the equilibrium condition for $\varepsilon$,
\begin{equation}
\tanh{\Delta\varepsilon\over\alpha r}+{\delta\over\bar w}=0,
\end{equation}
holds throughout the dynamics.
This is indeed consistent with the assumption
that $\varepsilon/r$ remains finite.
$\dot r$ is then given by
\begin{equation}
{1\over2}\dot r
\approx-\sqrt{2\alpha r\over\pi}e^{-(\Delta^2+\varepsilon^2)/2\alpha r}
\bar w\Bigl({r\over\Delta^2}\Bigr)
\Bigl[1-\Bigl({\delta\over\bar w}\Bigr)^2\Bigr]^{1/2}<0.
\end{equation}
This shows that the state $(m,\,r)=(\theta,\,0)$ is actually
a stable point of the dynamics described by 2-DRT
under the condition $2\alpha<\theta<1$.

\end{document}